\documentclass[aps,pra,twocolumn,groupedaddress,amsmath,amssymb,letterpaper]{revtex4}
\ifx\pdfoutput\@undefined\usepackage[usenames,dvips]{color}
\else\usepackage[usenames,dvipsnames]{color}
\usepackage{graphicx}
\usepackage{bm}
\usepackage{amssymb}
\usepackage{hyperref}
\usepackage{color}

\usepackage{amsmath}

\newcommand{\be}{\begin{equation}}
\newcommand{\ee}{\end{equation}}

\begin{document}

\title{Parity anomaly laser}

\author{Daria A. Smirnova$^{1,2}$, Pramod Padmanabhan$^{3}$, and Daniel Leykam$^{3}$}
\affiliation{$^{1}$Nonlinear Physics Center, The Australian National University, Canberra ACT 2601, Australia \\
$^{2}$Institute of Applied Physics, Russian Academy of Sciences, Nizhny Novgorod 603950, Russia \\
$^{3}$Center for Theoretical Physics of Complex Systems, Institute for Basic Science (IBS), Daejeon 34126, Republic of Korea}

\begin{abstract}
We propose a novel supersymmetry-inspired scheme for achieving robust single mode lasing in arrays of coupled microcavities, based on factorizing a given array Hamiltonian into its ``supercharge'' partner array. Pumping a single sublattice of the partner array preferentially induces lasing of an unpaired zero mode. A chiral symmetry protects the zero mode similar to 1D topological arrays, but it need not be localized to domain walls or edges. We demonstrate single mode lasing over a wider parameter regime by designing the zero mode to have a uniform intensity profile.
\end{abstract}

\maketitle

The ``parity anomaly'' refers to unpaired, symmetry-breaking zero modes, appearing in systems with supersymmetry (SUSY) such as strained honeycomb lattices~\cite{semenoff1984,haldane1988,schomerus2013}. A few recent studies proposed high power single mode lasing based on these anomalous zero modes: Schomerus and Halpern analyzed a parity symmetry-breaking pump in a strained honeycomb lattice~\cite{schomerus2013}: the parity-breaking zero mode overlaps most strongly with the pump, resulting in a lower lasing threshold compared to the other parity-preserving modes. Hokmabadi et al. exploited mode pairing via SUSY~\cite{miri2013,elgainany2015,teimourpour2016,hokmabadi2018,walasik2018}, weakly coupling two SUSY partner lattices together and pumping one of them to achieve lasing in the zero mode. Finally, lasing in 1D arrays with topologically protected midgap states has been observed in a few recent experiments~\cite{schomerus2013b,pilozzi2016,polariton_lasing,zhao2017,parto2018,malzard2018,kruk2018}, where again pumping one of the subsystems (sublattices in this case) induces single mode lasing robust against coupling disorder. While these examples may seem unrelated, they are all ultimately mediated by the parity anomaly. 

A limitation of the strained lattice and topological midgap mode proposals is that they are based on large lattices described by band structures. For integrated laser applications, small arrays of several to dozens of individual elements are preferable because slow dynamical instabilities typically inhibit synchronization of larger arrays~\cite{longhi2018b,longhi2018}. Furthermore, lasing based on localized topological zero modes is not compatible with large mode volumes, suffering from mode competition when realistic (saturable) gain is taken into account~\cite{longhi_supermode}. While the SUSY approach by Hokmabadi et al. is compatible with small arrays, the coupling between the two subsystems breaks the exact SUSY, introducing the additional problem of how to optimally couple the two subsystems~\cite{walasik2018}. What is needed is a way to systematically apply the parity anomaly to small systems of coupled lasers.

Here we propose a general SUSY-inspired method to create unpaired zero modes for robust single mode lasing in arrays of coupled resonators. We demonstrate numerically improved performance of the resulting ``parity anomaly laser'' compared to topological edge mode-based lasers. In particular, in our approach the lasing mode is not localized, so it not only harness the gain most efficiently, but mode competition is also maximized, suppressing the onset of multimode lasing.

We will consider a generic tight-binding network of $M$ coupled resonators with gain modelled by semiclassical rate equations. In the absence of gain, the network is characterized by a Hermitian Hamiltonian matrix $H_A$, with diagonal entries describing resonator detunings with respect to some reference frequency and off-diagonal entries describing inter-resonator couplings. The eigenstates of $H_A$ form the optical supermodes of the array. Assuming the lowest eigenvalue $\omega_0$ of $H_A$ is non-degenerate, $H_A - \omega_0 + \epsilon$ is a positive definite Hermitian matrix for $\epsilon > 0$ and can be uniquely factorized via the Cholesky decomposition, $H_A - \omega_0 + \epsilon = LL^{\dagger}$, where $L$ is a lower triangular matrix with positive real diagonal elements. This factorization allows the introduction of SUSY via the supercharge operators
\be 
q = \left( \begin{array}{cc} 0 & L \\ 0 & 0 \end{array} \right), \quad q^{\dagger} = \left( \begin{array}{cc} 0 & 0 \\ L^{\dagger} & 0 \end{array} \right), \label{eq:q}
\ee
satisfying $q^2 = (q^{\dagger})^2 = 0$. The SUSY Hamiltonian $\mathcal{H} = \{q, q^{\dagger}\}$ commutes with the supercharge, $[\mathcal{H},q]=[\mathcal{H},q^{\dagger}]=0$, and is  block diagonal: $\mathcal{H} = \mathrm{diag}(LL^{\dagger},L^{\dagger}L) = \mathrm{diag}(H_A-\omega_0 +\epsilon,H_B)$, where $H_B = L^{\dagger}L$ is the superpartner. In other words, SUSY relates two decoupled subsystems $H_{A,B}$, analogous to bosonic and fermionic sectors of SUSY quantum mechanics. If $\epsilon \ne 0$, SUSY is spontaneously broken: the two blocks share exactly the same eigenvalues, even though they generally do not have the same symmetries~\cite{tomic1997,bai2006}. In the limit $\epsilon \rightarrow 0$, $H_A-\omega_0$ has an unpaired zero mode (i.e. $H_B$ does not have a partner zero mode) and SUSY is unbroken~\cite{witten,miri2014,longhi2013,yu2015,nakata2016}.

Previously, Refs.~\cite{elgainany2015,hokmabadi2018} suppressed multimode lasing by coupling the two subsystems $H_{A,B}$ together with strength $\kappa$ and pumping only $H_A$, so that all excited states have weaker gain due to their hybridization with $H_B$. But this is only effective for weak coupling $\kappa \lesssim \omega_1 - \omega_0$, where $\omega_1$ is the first excited state's energy, since $\mathcal{H}$ no longer commutes with the supercharge $q$.

Ref.~\cite{midya_arxiv} recently introduced a variation to this standard SUSY construction: since $q^2 = (q^{\dagger})^2 = 0$, $\mathcal{H}$ can also be factored as the square of a Dirac-like Hamiltonian $H_D = q + q^{\dagger} = \sqrt{\mathcal{H}}$. This new ``supercharge'' Hamiltonian $H_D$ describes a single array consisting of two sublattices (A and B), with no direct coupling between members of the same sublattice. This relation between SUSY in Schr\"odinger and Dirac Hamiltonians has been studied for 30 years~\cite{nogami1993,cooper1988,nieto2003}, but it has been applied to obtain topologically nontrivial lattice Hamiltonians only in the past year~\cite{arkinstall2017,barkhofen2018,kremer2018,midya_arxiv}. $H_D$ has a chiral symmetry: it anticommutes with $C = \mathrm{diag}(1_M,-1_M)$, where $1_M$ is the $M \times M$ identity matrix, implying $C$ maps any mode with energy $\omega$ to a partner with energy $-\omega$. Since the eigenmodes of $H_D$ cannot simultaneously be eigenmodes of $C$ they must excite both sublattices of $H_D$.

The only exception to this rule is any zero modes of $H_D$, which \emph{can} be simultaneously made eigenstates of $C$ and thus localized to a single sublattice, forming a parity anomaly. When only this sublattice is pumped the zero mode will have a lower threshold than all other modes, enabling robust single mode lasing. Crucially, this holds regardless of the details of the parent lattice Hamiltonian $H_A$; it is not limited to periodic lattices~\cite{schomerus2013b,polariton_lasing,zhao2017,parto2018,malzard2018,arkinstall2017,barkhofen2018}, and therefore one can optimize $H_A$ to improve the lasing performance of $H_D$. In particular, the topological mid-gap states of Refs.~\cite{polariton_lasing,zhao2017,parto2018} are localized, such that the gain saturation inevitably leads to multimode lasing. With our approach, we can design delocalized unpaired zero modes with a uniform intensity distribution, maximizing the mode competition and suppressing the onset of multimode lasing. 

The simplest example of a delocalized unpaired zero mode can be obtained from a ring of $M$ identical resonators, described by the parent Hamiltonian
\be 
H_A = \sum_{j=1}^M (\hat{e}_{j+1,j} + \hat{e}_{j,j+1}), \quad j+M \equiv j \label{eq:4site}
\ee
where matrices $\hat{e}_{j,k}$ describe coupling from site $k$ to $j$ (coupling strength $J$ is normalized to 1) and satisfy $\hat{e}_{j,k} \times \hat{e}_{p,q} = \delta_{k,p} \hat{e}_{j,q}$. Since $H_A$ has a discrete rotational symmetry, all its modes have a uniform intensity distribution. This, however, means that under uniform pumping all its modes will experience the same gain, inhibiting single mode lasing. 

The Cholesky decomposition of $H_A$ is (for even $M$)
\begin{align}
L^{\dagger} = \sum_{j=1}^{M-2} &\left( (j+1)\hat{e}_{j,j} +  j \hat{e}_{j,j+1} + (-1)^{j+1} \hat{e}_{j,M} \right)/\sqrt{j(j+1)} \nonumber \\ & \qquad + \sqrt{M/(M-1)}\left( \hat{e}_{M-1,M} + \hat{e}_{M-1,M-1} \right), \label{eq:cholesky}
\end{align}
from which the supercharge Hamiltonian $H_D = q + q^{\dagger}$ can be constructed using~\eqref{eq:q}. Formally $H_D$ has $2M$ sites, but the zero $M$th row of $L^{\dagger}$ implies one of the B sublattice sites is completely decoupled from all others, which can be eliminated to leave the zero mode on the A sublattice unpaired. Note that coupling strengths in~\eqref{eq:cholesky} are site-dependent and vary in sign, requiring a mixture of positive and negative couplings~\cite{keil2016,bandres2018}.

\begin{figure}
\centering
\includegraphics[width=\columnwidth]{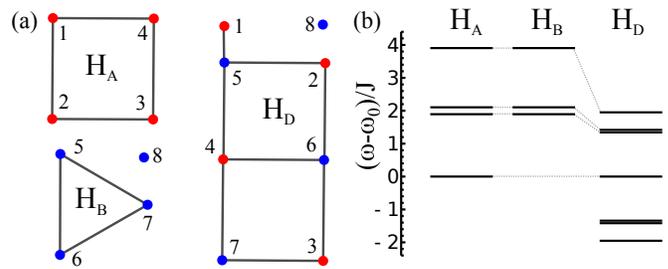}

\caption{(a) Schematic illustration of SUSY partner Hamiltonians $H_{A,B}$ and their supercharge Hamiltonian $H_D$ for an $M=4$ ring. Red (blue) sites correspond to A (B) sublattices of $H_D$. Note the intersite couplings (lines) are non-uniform (see text). (b) Corresponding energy spectra. The ground state of $H_A$ forms the unpaired zero mode of $H_D$.}

\label{fig:schematic}

\end{figure}

We now study in detail a supercharge array generated by an $M=4$ ring. Fig.~\ref{fig:schematic} illustrates the resulting SUSY and supercharge Hamiltonians, as well their energy spectra. The spectrum of $H_D$ is compressed (with normalized eigenvalues $\omega_n / J \rightarrow \pm \sqrt{\omega_n/J}$), increasing the frequency detuning between the zero mode and first excited state~\cite{midya_arxiv}. To describe a coupled laser array, we introduce gain and loss to $H_D$. We assume that only the A sublattice (hosting the unpaired zero mode) is pumped (inducing gain $g$), while the B sublattice has fixed loss $\gamma$. Close to the lasing threshold, where light intensities are low and gain saturation effects can be neglected, the emission spectrum under pulsed optical pumping is determined by the eigenvalues of the linear Hamiltonian $H_D + i \mathrm{diag}(g 1_4,-\gamma 1_4)$; lasing can occur in all modes whose eigenvalues have a positive imaginary part. 

\begin{figure}
\centering
\includegraphics[width=\columnwidth]{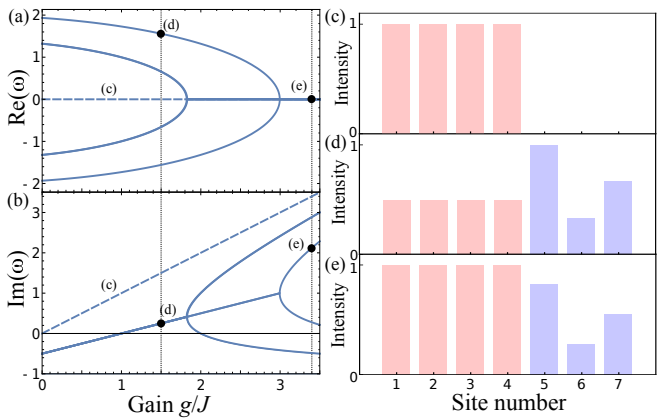}

\caption{Spectrum of $H_D$ when A sites are pumped with strength $g$ and B sites have fixed loss $\gamma = J$. (a,b) Real and imaginary parts of eigenvalues. Increasing the gain, excited states coalesce at non-Hermitian degeneracies at $\mathrm{Re}(\omega)=0$. (c) Unpaired zero mode profile [dashed lines in (a,b)]. (d) Representative excited state mode profile under weak pumping ($g/J=1.5$). Its power is distributed equally between the A (red) and B (blue) sublattices. (e) Excited state mode profile under strong pumping ($g/J=3.4$), localized more strongly to the A sublattice.}

\label{fig:eigenvalues}

\end{figure}

Fig.~\ref{fig:eigenvalues}(a,b) plots the energy eigenvalues of the array as a function of the gain $g$ for fixed loss $\gamma = J$. We observe behavior very similar to the recently demonstrated 1D topological laser arrays~\cite{polariton_lasing,zhao2017,parto2018}: the zero mode shown in Fig.~\ref{fig:eigenvalues}(c) most efficiently harnesses the gain ($\mathrm{Im}(\omega_0) = g$) and its threshold is independent of the loss on the B sublattice, since it is perfectly localized to the A sublattice. Other modes [see e.g. Fig.~\ref{fig:eigenvalues}(d,e)] initially have their power equally distributed between the two sublattices, reducing their slope efficiency and inducing a $\gamma$-dependent threshold. These modes only begin to localize to the pumped sites and compete with the zero mode when they reach the $\mathrm{Re}(\omega) = 0$ axis at a critical value of the gain-loss imbalance.

In contrast to the 1D topological laser arrays, here the zero mode uniformly excites all the pumped sites, so it can saturate the gain and maintain single mode operation over a wider range of parameters. To account for gain saturation we consider a class B laser model describing coupled resonators with embedded InGaAsP quantum wells~\cite{parto2018,longhi2018b,longhi2018},
\begin{subequations} \label{eq:evolution} 
\begin{align} 
i \partial_t \psi_n & = (H_D + H_{P}) \psi_n, \\
H_{P} \psi_n &= \frac{1}{2} \left( -\frac{1}{\tau_p} + \sigma (N_n - 1) \right)(i -\alpha) \psi_n, \\
\partial_t N_n &= R_n - \frac{N_n}{\tau_r} - \frac{2}{\tau_r} (N_n-1)|\psi_n|^2,
\end{align}
\end{subequations}
where $\psi_n$ and $N_n$ are the field amplitude and carrier density at the $n$th site, $\tau_p$ is the photon cavity lifetime, $\sigma$ is proportional to the differential gain, $\alpha$ is the linewidth enhancement factor, $R_n$ is the normalized (site-dependent) pumping rate, and $\tau_r$ is the carrier lifetime. For our numerical simulations, we take values representative of the pumped microring resonators considered in Refs.~\cite{longhi2018b,parto2018}: $\tau_p = 40$ ps, $\tau_s = 100 \tau_p$, $\sigma = 24/\tau_p$, $J = 3 / \tau_p$, and $\alpha = 3$, and assume the A and B sublattices are pumped with strength $R_A$ and $R_B$ respectively. 

The lasing intensity of the zero mode, $I_0 = \frac{1}{2}((R_A \tau_s - 1)\tau_p\sigma - 1)$, is obtained from~\eqref{eq:evolution} by assuming it saturates the gain on the A sublattice. It is identical to that of a single ring, being independent of the energy scale $J$ set by the inter-site couplings and the pump rate on the B sublattice. Nevertheless, its stability is sensitive to these parameters. We perform the linear stability analysis by linearizing~\eqref{eq:evolution} about the zero mode solution (see e.g. Ref.~\cite{longhi2018}). The zero mode is linearly stable if none of its perturbation eigenvalues have a positive real part.

\begin{figure}

\centering
\includegraphics[width=0.5\columnwidth]{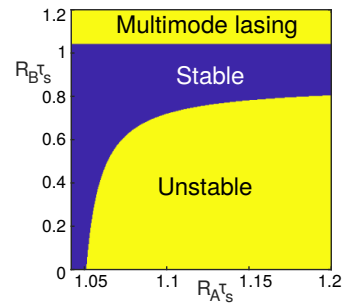}

\caption{Stability diagram of the zero mode as a function of the A $(R_A)$ and B $(R_B)$ sublattice pump strengths. The lower limit of $R_A$ is the lasing threshold.}

\label{fig:stability}

\end{figure}

Fig.~\ref{fig:stability} plots the stability as a function of the two pump rates $R_{A,B}$, revealing a large parametric domain of stable single mode lasing. When the B pump rate $R_B$ is low, the effective coupling between the pumped sites is weak due to the high losses in the B sites. In this weakly-coupled regime, when the zero mode first becomes unstable the system converges to a stable asymmetric mode, predominantly localized to the A sublattice. At even stronger pump rates, this asymmetric mode undergoes a Hopf bifurcation into a limit cycle, before chaotic dynamics emerge~\cite{photonic_dimer}. Meanwhile, the upper multimode lasing region in Fig.~\ref{fig:stability} corresponds to the pump $R_B$ exceeding losses at the B sites. In this case lasing occurs at multiple frequencies with splitting determined by the inter-site coupling $J$.

To validate the above linear stability analysis, we also solved~\eqref{eq:evolution} numerically using the 4th order Runge-Kutta method, starting from random noise in the optical field $\psi_n$ and equilibrium carrier densities $N_n(0) = R_n \tau_s$. We assume pump rates identical to those used in the recent experiments of Ref.~\cite{parto2018}: $R_A = 1.07/\tau_s$ (above threshold) and $R_B = 1.03/\tau_s$ (slightly below threshold). This corresponds to initial (small signal) gain and loss rates of $g \approx 0.07J$ and $\gamma \approx g/10$ respectively.

\begin{figure}

\centering
\includegraphics[width=\columnwidth]{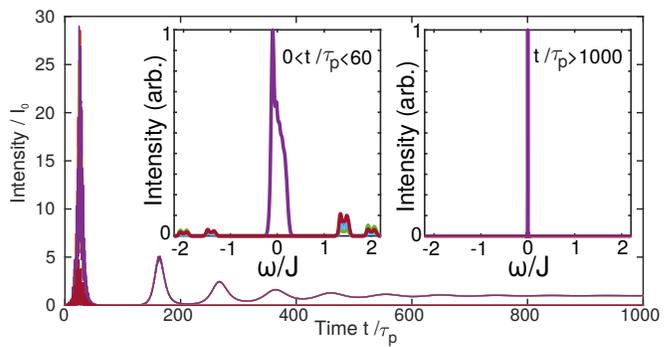}

\caption{Switch-on dynamics of the laser, revealing suppression of all excited states due to gain saturation and stable lasing in the unpaired zero mode. Insets show Fourier spectra during (left) and after (right) the initial transient.}

\label{fig:laser1}

\end{figure}

Fig.~\ref{fig:laser1} illustrates the resulting dynamics. Initially all linear modes are above threshold and start to grow, such that the emission spectrum measured under pulsed optical pumping would reveal multiple peaks corresponding to the linear modes of $H_D$ (see left inset in Fig.~\ref{fig:laser1}). At longer times, however (e.g., under continuous electrical pumping), the gain saturation plays a critical role. The zero mode grows the fastest, and since it uniformly excites all pumped sites it is able to completely saturate the gain and suppress all other modes. Thus, after an initial transient the system converges to the stable unpaired zero mode lasing state. Similar to the midgap states of the Su-Schrieffer-Heeger chain~\cite{schomerus2013b,polariton_lasing,zhao2017,parto2018,malzard2018}, the zero mode is protected by the chiral symmetry of $H_D$ and we checked that this stable operation persists even under moderate coupling disorder. 

\begin{figure}
\centering
\includegraphics[width=\columnwidth]{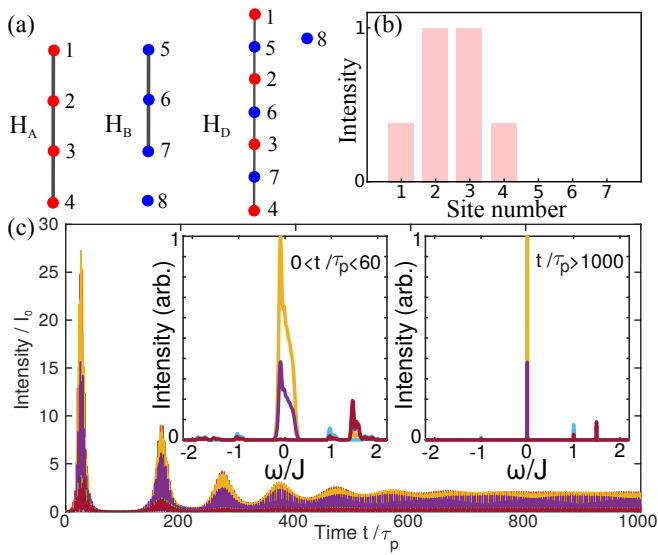}

\caption{Parity anomaly laser derived from a 1D chain. (a) $M=4$ site chain, its SUSY partner $H_B$, and supercharge Hamiltonian $H_D$. (b) Unpaired zero mode with non-uniform intensity on majority sublattice. (c) Laser switch-on dynamics, revealing persistent multimode lasing because the zero mode cannot saturate the gain. Insets: Fourier spectra during (left) and after (right) initial transient.}

\label{fig:laser2}

\end{figure}

To demonstrate the importance of the uniform mode profile in achieving the stable single mode lasing, Fig.~\ref{fig:laser2} compares the performance of a parity anomaly laser model generated from an \emph{open} chain. In this case, the parent Hamiltonian lacks any discrete rotational symmetry and the zero mode does not uniformly excite all the pumped sites (similar to the case of the 1D topological edge state lasers). Hence the zero mode cannot saturate the gain by itself and lasing in excited states persists.

\begin{figure}

\includegraphics[width=\columnwidth]{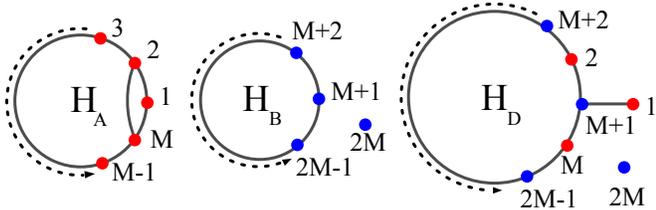}

\caption{Schematic illustrating SUSY partners $H_{A,B}$ generating a scalable supercharge laser array $H_D$ with uniform zero mode and purely local coupling for arbitrary array size $M$. Note the intersite couplings (lines) are non-uniform (see text).}

\label{fig6}

\end{figure}

Larger array sizes $M$ are desirable for applications such as increasing the output power of integrated semiconductor laser arrays. Unfortunately, with increasing $M$ the supercharge Hamiltonian generated from the simple ring Cholesky decomposition~\eqref{eq:cholesky} becomes highly nonlocal, involving a central ``hub'' site coupled to many others. However, our approach can also be used to generate more physically reasonable large $M$ supercharge arrays by directly optimizing the matrix $L^{\dagger}$. For example, if we assume only nearest neighbor couplings are nonzero, the most general form of $L^{\dagger}$ supporting an unpaired zero mode with uniform intensity is
\begin{align} \label{2ndL}
L^{\dagger} = \kappa_1 \hat{e}_{1,2} + \kappa_2 \hat{e}_{1,M} + \kappa_3 \hat{e}_{1,1} + \sum_{j=2}^{M-1} c_j \left( \hat{e}_{j,j} + \hat{e}_{j,j+1} \right),
\end{align}
where $\kappa_3 = \kappa_1 + (-1)^M \kappa_2$ and other couplings $c_j$ and $\kappa_{1,2}$ are free parameters. The resulting SUSY and supercharge networks are illustrated in Fig.~\ref{fig6}. Here the uniform intensity is enabled by the paired coupling terms $c_j$, while the additional defect site side-coupled by $\kappa_1$ allows the zero mode to be unpaired. In this case the SUSY Hamiltonians $H_{A,B}$ both form closed rings with local couplings. Interestingly, the zero mode of $H_A$ has a uniform intensity despite the couplings, site detunings, and excited states all being inhomogeneous.

In summary, we propose a supersymmetry-inspired method to achieve stable single mode lasing in coupled microcavity networks, based on designing ``supercharge'' resonator networks hosting unpaired zero modes~\cite{midya_arxiv}. The advantage of our scheme is that the symmetry-protected zero mode is not localized and is therefore able to completely saturate the gain, preserving single mode operation above the thresholds observed in Refs.~\cite{polariton_lasing,zhao2017,parto2018}. An interesting future direction would be to generalize this scheme to genuinely two-dimensional arrays~\cite{bahari2017,bandres2018,klembt2018}. Experimentally, these ideas can be readily tested using coupled ring resonator lattices with optical gain provided by embedded quantum wells, where a conventional SUSY laser was recently demonstrated~\cite{hokmabadi2018}. This mechanism for achieving stable phase locking of small laser arrays may be useful for increasing the power output of semiconductor lasers, which is particularly important when material constraints such as the optical damage threshold limit the output power from individual lasers.

This work was supported by the Institute for Basic Science (IBS-R024-Y1), the Russian Foundation for Basic Research (Grant 18-02-00381), and the Australian Research Council. We thank Vassilios Kovanis and Sunkyu Yu for enlightening discussions.

\end{document}